\documentclass[sigconf]{acmart}

\def\BibTeX{{\rm B\kern-.05em{\sc i\kern-.025em b}\kern-.08emT\kern-.1667em\lower.7ex\hbox{E}\kern-.125emX}}
    
\copyrightyear{2019}
\acmYear{2019} 
\setcopyright{iw3c2w3}
\acmConference[WWW '19 Companion]{Companion Proceedings of the 2019 World Wide Web Conference}{May 13--17, 2019}{San Francisco, CA, USA}

\acmBooktitle{Companion Proceedings of the 2019 World Wide Web Conference (WWW '19 Companion), May 13--17, 2019, San Francisco, CA, USA}
\acmPrice{}
\acmDOI{10.1145/3308560.XXXXXX}
\acmISBN{978-1-4503-6675-5/19/05}

\usepackage{balance}
\begin{document}

% The "title" command has an optional parameter, allowing the author to define a "short title" to be used in page headers.
\title{Human Values and Attitudes towards Vaccination in Social Media}

%
% The "author" command and its associated commands are used to define the authors and their affiliations.
% Of note is the shared affiliation of the first two authors, and the "authornote" and "authornotemark" commands
% used to denote shared contribution to the research.

\author{Kyriaki Kalimeri}
\affiliation{Data Science Laboratory,  ISI Foundation,  Turin,  Italy}
  
\author{Mariano G. Beir\'{o} }
%\email{larst@affiliation.org}
\affiliation{Universidad de Buenos Aires, Facultad de Ingenier\'{i}a,\\ INTECIN (CONICET), CABA, Argentina}

\author{Alessandra Urbinati}
%\email{larst@affiliation.org}
\affiliation{Data Science Laboratory,  ISI Foundation,  Turin,  Italy}

\author{Andrea Bonanomi}
\affiliation{Universit\`{a} Cattolica del Sacro Cuore (UNICATT), Milan, Italy}

\author{Alessandro Rosina}
\affiliation{Universit\`{a} Cattolica del Sacro Cuore (UNICATT), Milan, Italy}

\author{Ciro Cattuto}
\affiliation{Data Science Laboratory,  ISI Foundation,  Turin,  Italy}

%
% By default, the full list of authors will be used in the page headers. Often, this list is too long, and will overlap
% other information printed in the page headers. This command allows the author to define a more concise list
% of authors' names for this purpose.
\renewcommand{\shortauthors}{Kalimeri, et al.}

%
% The abstract is a short summary of the work to be presented in the article.
\begin{abstract}

Psychological, political, cultural, and even societal factors are entangled in the reasoning and decision-making process towards vaccination, rendering vaccine hesitancy a complex issue.
Here, administering a series of surveys via a Facebook-hosted application, we study the worldviews of people that 
``Liked'' supportive or vaccine resilient Facebook Pages. 
In particular, we assess differences in political viewpoints, moral values, personality traits, and general interests, 
finding that those sceptical about vaccination, appear to trust less the government, are less agreeable, while they are emphasising more on anti-authoritarian  values. 
Exploring the differences in moral narratives as expressed in the linguistic descriptions of the Facebook Pages, we see that pages that defend vaccines prioritise the value of the family while the vaccine hesitancy pages are focusing on the value of freedom.   
Finally, creating embeddings based on the health-related likes on Facebook Pages, 
we explore common, latent interests of vaccine-hesitant people, showing a strong preference for natural cures.
This exploratory analysis aims at exploring the potentials of a  social media platform to act as a sensing tool, providing researchers and policymakers with insights drawn from the digital traces, that can help design communication campaigns that build confidence, based on the values that also appeal to the socio-moral criteria of people.

\end{abstract}



\maketitle

\section{Introduction}

%{\color{red} ideas for improvement\\
%- predict political values from likes\\
%- put results of prediction in text\\
%- do literature review on social media usage for vaccine hesitancy\\
%- think of a nicer title
%}
Over the last decades, vaccines have saved countless lives and are undoubtedly the most successful and cost-effective intervention to improve health, both in an individual and at a community level. 
Despite their effectiveness, vaccine hesitancy is becoming an emerging issue both in high- and low-income countries \cite{larson2011addressing,salmon2015vaccine}. 
According to the W.H.O. \cite{domek2018measuring}, to the present day, ``no uniform, global metric for quantifying vaccine hesitancy currently exists''.
Approximately 40\% of parents in the United States may delay or refuse vaccinations for their children~\cite{smith2011parental}.

A growing number of people use the Internet and social media to obtain information about health-related issues, including information about vaccines.
Nowadays, social media platforms, such as Twitter and Facebook, are becoming increasingly more popular sources of health information  \cite{kennedy2011confidence,capurro2018measles} despite their content being often subject to popularity dynamics. 
As a recent study shows, the consumption of vaccine-related content on Facebook is dominated by the ``echo chamber'' effect \cite{schmidt2018polarization}.
%With the proliferation of conflicting information regarding vaccines, obtaining health information can create an even more confusing context \cite{betsch2012opportunities}. 

Reasoning on the evidence around vaccination is hardly ever the outcome of a fact-driven analysis.
Vaccine hesitancy is a complex issue with numerous underlying factors influencing the decision-making process~\cite{schmid2017barriers}.
The determinants of vaccination endorsement may range from socio-demographic to psychological, emotional, and cultural factors \cite{dube2015strategies,dube2016nature,greenberg2017vaccine,shapiro2018vaccine}. 
Political opinions, too, influence the willingness to trust advice received by governments or evidence-based medical models \cite{kata2010postmodern}.
%Due to the multi-facet nature of the problem the is still no effective communication strategy \cite{dube2015strategies}.
Since taking a stance regarding vaccination is a broader psychological and moral decision-making process, opinions are unlikely to change by appeals to reasoning or evidence alone \cite{browne2015going}. 
Hence, the design of effective communication interventions requires an in-depth understanding of all the determinants of vaccine hesitancy of the specific population \cite{attwell2018recent}.

%\cite{Amin2017}
Psychological attributes and worldviews are reflected in a wide range of digital traces, indicatively on smartphone data~\cite{Xu2016,Kalimeri2019predicting}, Twitter~\cite{Quercia2011}, and Facebook~\cite{Youyou2015}.
Here, we engaged a large cohort in Italy ($N=34,200$), using a Facebook application. Participants of this cohort were invited to complete surveys on a range of topics including psychological, moral, and political views. 
They allowed us to access their Likes on Facebook Pages;
assuming that ``liking'' a Page with positive or negative attitudes towards vaccination reflect the opinion of the person on the topic.
We compared the psychological and moral profile of 
people with a positive or negative attitude towards vaccination,
presenting interesting insights on how the two groups differ.
Moreover, addressing linguistically the moral narratives provided on the description of each page, we give an insight into the values that are more emphasised on the two communities.
Finally, we create a network embedding based on the Pages ``Liked'' by our participants to explore common interests these people may share.
%giambi2018parental

The broader goal of this study is to investigate whether Facebook can be employed as a sensing tool, as a proxy to the narratives that determine vaccination endorsement. Digital data from social media can assist researchers and policymakers in understanding better the psychological profile, moral worldviews, but also the discourses underlying anti-vaccinationism on a larger scale.
Such insights may inform interventions that take into consideration the cognitive, moral,  psychological, and political values that are more likely to appeal to the vaccine-hesitant people.
  
%\cite{larson2011addressing}
%- perhaps I can do a prediction of our social Q from pages? - MARIANO what do you think?
%-next on... conslusions and abstract.

\section{Data Collection and Methods}
% cite demography paper (kalimeri2019evaluation)
% and unemployment...eh... not yet out...

\subsection{Related Work and Theoretical Background}

Previous research on the determinants of attitude formation towards vaccination suggests that there are broader psychological, political, cultural, or even societal factors that may contribute to negative vaccination attitudes \cite{bean2011emerging,kata2010postmodern,yaqub2014attitudes,schmid2017barriers}.
More specifically, vaccination scepticism has been related to  unwillingness to engage with the scientific evidence\cite{browne2015going}, 
an alignment with alternative/complementary or holistic health \cite{kata2010postmodern}, as well as
spiritual and religious identities~\cite{browne2015going,kata2010postmodern},
anti-authoritarian worldviews\cite{browne2015going},
conspiracy ideation \cite{jolley2014effects},
trust and political attitudes \cite{yaqub2014attitudes}.

In this study, we operationalise the political viewpoints, moral values, personality traits, as well as cultural elements of both supporters and refuters of vaccination via self-reported questionnaires, aiming to provide a holistic view of the beliefs of the two communities.

\subsubsection*{\textbf{Political Values}}
\label{sec:politics}
Our moral, psychological, and ethical values mediate relations to opinions and attitudes towards major societal issues.
Inspired by the 41-item inventory proposed by Schwartz et al.\cite{schwartz1992universals} and the updated version of Barnea et al.~\cite{barnea1998values} we included a 15-item survey to the participants of \textit{Likeyouth} since the full version of the original inventories would be too long for an administered study on Facebook. 
The questions included regarding essential issues of the society, often arguments of long political debates, such for example opinions on immigration, fair treatment of individuals, trust in the Government and the European Union. The participants were asked to rate the following items, on a 5-point Likert scale, how much they agreed or not with the following phrases.

%{\color{red} check the translation...OK I found the source
%https://www.researchgate.net/profile/Shalom_Schwartz/publication/229789582_Basic_Personal_Values_Core_Political_Values_and_Voting_A_Longitudinal_Analysis/links/59dfb402458515371600dd12/Basic-Personal-Values-Core-Political-Values-and-Voting-A-Longitudinal-Analysis.pdf
%}

\begin{itemize}
    \item \textit{Q1}: It is extremely important to defend our traditional religious and moral values.
    \item \textit{Q2}:  It would be a good idea to privatise all of the public enterprises.
    \item \textit{Q3}: People who come to live here from other countries generally make our country a better place to live.
    \item \textit{Q4}:  If people were treated more equally in this country, we would have many fewer problems.
    \item \textit{Q5}: I trust the President of the Republic.
    \item \textit{Q6}: Being rich is important to me. I want to have a lot of money and buy expensive things.
    \item \textit{Q7}: I believe every person in the world should be treated in the same way. I believe that everyone should have the same opportunities in life.
    \item \textit{Q8}: I trust the national government.
    \item \textit{Q9}: Newer lifestyles are contributing to the breakdown of our society.
    \item \textit{Q10}: The less the government gets involved with business and the economy, the better off this country will be.
    \item \textit{Q11}: Foreigners that come to live in our country threaten the harmony of our society.
    \item \textit{Q12}: It is extremely important to respect the freedom of individuals to be and believe whatever they want.
    \item \textit{Q13}: I trust the European Union.
    \item \textit{Q14}: Being successful is important to me. I like to impress people.
    \item \textit{Q15}: I strongly believe that the state needs to be always aware of the threads both internal and external.
\end{itemize}
       
\subsubsection*{\textbf{Moral Foundations}}
The moral values were assessed via the Moral Foundations Theory (MFT); a psychologically validated questionnaire, which focuses on the explanation of the psychological basis of morality, its origins, development, and cultural variations, 
and identifies the following five moral foundations \cite{Haidt2007,Haidt2004}:
\begin{itemize}
   \item \textbf{Care/Harm}: basic concerns for the suffering of others, including virtues of caring and compassion.
   \item \textbf{Fairness/Cheating}: concerns about unfair treatment, inequality, and more abstract notions of justice,
   \item \textbf{Loyalty/Betrayal}: concerns related to obligations of group membership, such as loyalty, self-sacrifice and vigilance against betrayal,
   \item \textbf{Authority/Subversion}: concerns related to social order and the obligations of hierarchical relationships
such as obedience, respect, and proper role fulfilment.
   \item \textbf{Purity/Degradation}: concerns about physical and spiritual contagion, including virtues of chastity, wholesomeness and control of desires.
\end{itemize}
These foundations collapse into two superior foundations \cite{Haidt2007}; the \textbf{individualising}, which asserts that the basic constructs of society are the individuals and hence focuses on their protection and fair treatment, and the \textbf{binding} foundation, based on the respect of leadership and traditions.
%The assessment of personality and morality traits is of major importance since these attributes are often fundamental explanatory factors tied to many sociological phenomena directly tied to demographic issues \cite{Bi2013,Kalimeri2017}.

\subsubsection*{\textbf{Personality Traits}}
The personality was assessed via the Big5 personality traits model \cite{gosling2003very,Costa1992}; a well-established theory, which characterises personality based on the following five dimensions with universal validity~\cite{schmitt2007geographic}:
\begin{itemize}
    \item \textbf{Openness to experiences}: inventive/ curious vs. consistent/ cautious,
    \item \textbf{Conscientiousness}: efficient/ organised vs. easy-going / careless,
    \item \textbf{Extraversion}: outgoing/ energetic vs. solitary/ reserved,
    \item \textbf{Agreeableness}: friendly/ compassionate vs. analytical/ detached,
    \item \textbf{Neuroticism}: sensitive/nervous vs. secure/ confident. 
\end{itemize}

\subsubsection*{\textbf{Interests \& Hobbies}}
Additionally, to the psychologically validated questionnaires, such as the ones mentioned above, the application also contains surveys regarding interests and opinions in general.  In one of the surveys, the participants were asked to rate their interest, on a 5-point Likert scale, in the following categories: Travel, Sport, Science, Food, Culture, Nature, Society and Politics, Education, Health, Hobbies, Business, Shopping.

\subsection{Data Collection}

Psychological, moral, and political opinion questionnaires were administered online, on  \emph{Likeyouth}\footnote{\href{http://likeyouth.org}{http://likeyouth.org}}, a Facebook-hosted application. 
The Facebook platform is shown to be a valid scientific tool for administering questionnaires, with eligible population, self-selection, and behavioural biases~\cite{Kalimeri2019}.

This application acts as an innovative data-collection tool, gathering apart from the self-reported psychological and opinion assessments previously described also the  ``Likes'' on the Facebook Pages.
A consent form was obtained, from all participants, regarding a privacy agreement which they declare to accept upon registration. 
Once participants enter the application, they are asked to fill in their profile including basic demographic information such as gender, age, employment status, and region of residence.
The application is mainly deployed in Italy; it was initially launched in March 2016~\cite{Bonanomi2017}, while we downloaded the data on September 2018. 
Overall 34,200 participants (11,315 female), of average age 33 years old, entered the application. This sample is representative of the geographical distribution of the Italian population (Spearman correlation of $0.88$ with the population at province level in Italy).

Among the Pages ``Liked'' by our participants, we searched for those who contained the stem {vacc*} in the Page name. These were then manually annotated according to their relevance to the topic of vaccination, avoiding irrelevant pages such for instance the musical band ``The Vaccines''. 
The relevant pages to the topic of vaccination were further annotated according to their content in supportive (hereafter, PV, for brevity) or contrary (hereafter, AV) to vaccination. 
%The satirical pages were kept out of the analysis for the scope of this study.
In total, we result with 44 PV pages, with an average of 11,122 pages ``likes'', 53 AV pages, with 20,238 page ``likes'' on average, and 13 satyric pages (all PV), with 23,823 ``likes'' on average. The satirical pages were aggregated to the PV pages since they were all supportive of the vaccination.

%%%
%{\color{red} perhaps add here a subsection with the statistical/wordcloud analysis?}
 
\subsection{Network Embedding Definitions}

We aimed at exploring the general interests of people with conflicting views on vaccination, as expressed through their ``likes'' on Facebook Pages.
In doing so, we propose a fully unsupervised approach based on network embeddings that exploits the intrinsic properties of the network without any a priori information about the participants. 

We consider the network $G = (V,\varpi)$, where $V$ is a set of nodes, and $\varpi: V  \times V \rightarrow  \mathbb{N}$ is a function defining for each pair of nodes $i,j \in V$ the weight of edge $(i,j)$, that is $w_{ij}$.
Since, in this study, we are interested in getting insights regarding the participants' interests, we modelled as nodes of the network our participants, while the co-occurrence of ``likes'' on pages are employed to create the links between the participants. 
Hence, if the participant \textit{i} and participant \textit{j} share - have ``Liked'' - at least one page in common,  then they share a link, while the weight of the link, defined as $\varpi$, represents the exact number of the common pages that the two endpoints share.

The network embeddings are created employing 
the \textit{node2vec} implementation~\footnote{Link to the reference implementation: \url{https://snap.stanford.edu/node2vec/}}, a well-established and efficient algorithm \cite{GroverL16,goyal2018graph}.
Mapping of nodes to a feature space of lower dimension maximises the likelihood of preserving network neighbourhoods.
The neighbourhood exploration is based on a second-order random walk, with two parameters, namely \textit{p} and \textit{q} that guide the walk. Parameter \textit{p} controls the likelihood of immediately revisiting a node in the walk, while  \textit{q} allows differentiating the search. If $q>1$ the walk retains a local view of the neighbourhood, otherwise the walk is more inclined to visiting nodes which are further away from the current one. Each walk, that has just traversed the link $(t,i)$ decide the next step according to the following non-normalised transition probabilities, evaluated on each edges $(i,j)$ leaving from $i$, $\pi_{ij}=\alpha_{pq}(t,j)\cdot w_{ij}$, where $w_{ij}$ is the link's weight, 
\begin{equation}
\alpha_{pq}(t,j) = 
  \begin{cases}
      \frac{1}{p} & \text{if $d_{tj}=0$}\\
      1 & \text{if $d_{tj}=1$}\\
      \frac{1}{q} & \text{if $d_{tj}=2$}\\
      \end{cases} 
\end{equation}
and $d_{tj}$ is the shortest path distance between nodes $t$ and $j$.\\
In this way, nodes belonging to the same neighborhood preserve a ``structural'' equivalence or homophily. Real networks usually exhibit a mixture of both.

\section{Results and Discussion}

In our dataset, we have 1777 unique participants in total who expressed an interest in vaccine controversies; 1907 who followed PV pages, while 113 participants followed AV pages. Of course, a small fraction of the participants followed both PV and AV pages. In the following sections, we discuss the differences and similarities emerged from the comparison of the two populations.
Note that due to the spontaneous nature of Likeyouth application, not all participants completed all the questionnaires.

\subsubsection*{\textbf{Demographics}}
Demographic information is obtained via self-reporting once the participants enter the application. Data on employment status and residence were not sufficient for concluding the population of the two cohorts; gender information, on the other hand, was available for 1739 participants.
Proportionally we have slightly more males in the PV group (62\% males and 36\% females), with respect to the AV group (52\% males and 47\% females) (p-value$<0.05$ on Fisher's exact test).
Such observation was reported in other scientific studies, for instance, ~\cite{gilkey2013forgone, ekos2018}.
Scientists systematically try to uncover the reasons why mothers appear to be more sceptical in whether to accept a vaccine or not.
It's a complex and multi-facet issue, it seems that many mothers consider their child's immune system to be ``unique''~\cite{cassell2006cultural} while their decision-making process ``encompassed different factors such as social norms, past experiences, emotions, values, social network influences, and other day-to-day concerns about their child's health and well-being''~\cite{dube2016nature}.

\subsubsection*{\textbf{Political Values}}

The political values survey was completed by 1232 participants who also expressed interest in vaccine controversies. Comparing the PV and AV cohorts using the Mann-Whitney U test, we found a few statistically significant differences. 
AV seem to trust less the governmental norms, including the President of the Republic ($Q5$, p-value $< 0.001$), the national government ($Q8$, p-value $< 0.001$), as well as the European Union ($Q13$, p-value $< 0.001)$. 
These results are coherent also with the findings of Browne et al.~\cite{browne2015going}, unwillingness to trust information delivered by conventional authority sources is a predictor of negative attitudes to vaccination.
Further, there seems to be a propensity of the AV cohort to defend the traditional religious and moral values ($Q1$), p-value $= 0.011$, and also argue that the newer lifestyles contribute to the decline of our society ($Q9$), p-value $< 0.001$.
This finding contrast studies carried out in the US, where people with negative views towards vaccination seem to value more the notion of freedom, concerning personal expression, and religious beliefs, to the vaccine supporters~\cite{Amin2017}.
%We note - even if the statistical significance test does not support the result - that PV participants appear to have a more favourable opinion regarding the immigrants and their assimilation in the society ($Q3$). 
Table~\ref{tab_social}, reports the differences in core political values between the two cohorts.

\begin{table}
\begin{tabular}{lccc}
\hline
Attribute & PV median & AV median & p-value \\
\hline
\textbf{\textit{Q1}} & 2.0 & 3.0 & 0.011\\
\textit{Q2} & 2.0 & 2.0 & 0.395 \\
\textit{Q3} & 4.0 & 3.0 & 0.160\\
\textit{Q4} & 5.0 & 5.0 & 0.170\\
\textbf{\textit{Q5}} & 4.0 & 3.0 & $<$0.0001\\
\textit{Q6} &3.0 & 3.0 & 0.214\\
\textit{Q7} & 5.0 & 5.0& 0.498\\
\textbf{\textit{Q8}} & 3.0 & 2.0 & $<$0.0001 \\
\textbf{\textit{Q9}} & 2.0 & 3.0 & 0.001\\
\textit{Q10} & 3.0& 3.0 & 0.478\\
\textit{Q11} & 2.0 & 2.0 & 0.176\\
\textit{Q12} & 5.0 & 4.0 & 0.121\\
\textbf{\textit{Q13}} & 4.0 & 3.0 & $<$0.0001 \\
\textbf{\textit{Q14}} & 3.0 & 3.0 & 0.010\\
\textit{Q15} & 4.0 & 4.0 & 0.836\\
\hline
\end{tabular}
\caption{Median values and p-values for the Mann-Whitney tests on the difference in opinions in social issues between PV and AV.}
\label{tab_social}
\end{table}

\subsubsection*{\textbf{Moral Values}}
%MFT_proVAX_Only_vs_NoVAX_Only_
Overall, 128 participants interested in vaccine controversy page filled in the moral foundations questionnaire. %($N_PV$ = 124, $N_AV$ = 14). 
To compare the two cohorts, i.e. the population that supports vaccination practices, PV, to the one that is more sceptical about it, AV, we employed the non-parametric Mann-Whitney U test \cite{mann1947test}.
Due to the limited size of the cohort, none of the results is statistically significant. 
%Despite this, we note that AV participants value care more, authority less, and they are much more individualists. 
%Amin et al. \cite{Amin2017}, in their recent study, found a significant association between the moral values of purity and authority with vaccine hesitancy. People sceptical about vaccination were less authoritarian and strongly emphasised purity. 
%AV supporters also scored higher in care, giving more emphasis to the suffering of others.
%The portrait of the moral profiles of the two groups can be summarised in the two superior foundations;
%AV appears to be more individualists, hence focusing more on the protection and fair treatment of the individuals. 
%In contrast, PV participants appear to emphasise more the social binding foundation, putting the society as a priority over the individual, which is in line with the current literature on the topic~\cite{Amin2017}.

\begin{table}
\begin{tabular}{lccc}
\hline
Attribute & PV median & AV median & p-value \\
\hline
\textbf{Care} & 19.0 & 21.0 & 0.09 \\
Fairness & 21.0 & 22.0 & 0.42 \\
Loyalty & 15.0 & 18.0 & 0.899 \\
Authority & 14.0 & 13.0 & 0.667 \\
Purity & 14.0 & 14.0 & 0.390 \\
\textbf{Individualism} & 40.0 & 43.0 & 0.213 \\
\textbf{Social binding} & 43.0 & 39.0 & 0.921 \\
\hline
\end{tabular}
\caption{\label{tab_moral}Median values and p-values for the Mann-Whitney tests on the difference in moral values between PV and AV participants.}
\end{table}

\subsubsection*{\textbf{Personality Traits}}

Only a small fraction of the participants completed the personality traits questionnaire, 116 participants in total. 
Due to the small sample size, the only statistically significant result was the tendency of the PV participants to score higher in the agreeableness trait (p-value$<0.05$). Table~\ref{tab_big5} summarised the findings of all the five traits. 
Low agreeableness scores have been previously related to vaccine hesitancy \cite{lee2017personality}.
%Also, higher scores in openness to experience trait too have been associated to lower confidence about vaccine safety \cite{browne2015going,lee2017personality}, in our study this pattern emerges, however, the statistical significance test does not support it.

\begin{table}
\begin{tabular}{lccc}
\hline
Attribute & PV median & AV median & p-value \\
\hline
Extraversion & 8.0 & 8.0 & 0.793 \\
\textbf{Agreeableness} & 10.0 & 8.0 & 0.025 \\
\textbf{Conscientiousness} & 9.0 & 11.0 & 0.157 \\
Openness & 8.0 & 9.0 & 0.315 \\
Neuroticism & 9.5 & 9.5 & 0.924 \\
\hline
\end{tabular}
\caption{\label{tab_big5}Median values and p-values for the Mann-Whitney tests on the difference in personality traits between PV and AV participants.}
\end{table}

\subsubsection*{\textbf{Interests and Hobbies}}
Self-reported information quantifying the broader interest in general topics, for instance ``science'',
was available for 153 participants. Table~\ref{tab_fb_interests} reports the Mann-Whitney U test on the two cohorts pointed out that AV participants are more into hobbies in general, but interestingly, are significantly more interested in sports and health-related topics. 
Such indication points out that since these individuals are more interested in health, will actively look for information about vaccination; the challenge lies in the documentation they will be more likely to receive \cite{schmidt2018polarization,waszak2018spread} but also their predisposition in interpreting what they read \cite{kata2010postmodern,betsch2012opportunities,capurro2018measles}.

\begin{table}
\begin{tabular}{lccc}
\hline
Attribute & PV median & AV median & p-value \\
\hline
Travel & 3.0 & 4.0 & 0.561 \\
\textbf{Sport} & 3.0 & 4.0 & 0.027 \\
Science & 4.0 & 3.0 & 0.356 \\
Food & 4.0 & 4.0 & 0.522 \\
Culture & 4.0 & 5.0 & 0.744 \\
Nature & 4.0 & 4.0 & 0.426 \\
Society \& Politics & 4 & 4 & 0.976 \\
Education & 4 & 4 & 0.749 \\
\textbf{Health} & 3.0 & 4.0 & 0.023 \\
\textbf{Hobbies} & 3.0 & 4.0 & 0.049 \\
Business & 4 & 4 & 0.473 \\
Shopping & 2 & 2 & 0.423 \\
\hline
\end{tabular}
\caption{Median values and p-values for the Mann-Whitney U test on the difference in topics of interest between PV and AV participants.}
\label{tab_fb_interests}
\end{table}

\subsection{Moral Values in Facebook Pages}

Having assessed the emerging differences in the political, moral, and psychological worldviews of individuals with demonstrated positive or negative attitudes towards vaccination, we place the focal point on the understanding of the content presented in the PV and AV pages from a linguistic perspective.

\begin{figure*}
\includegraphics[height=6.5cm]{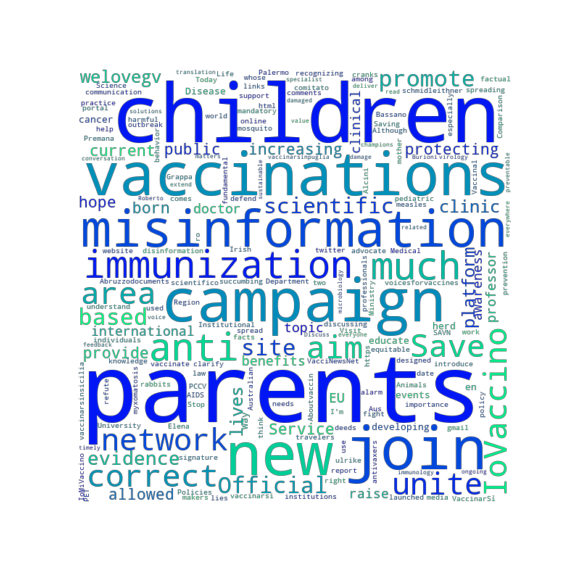}
\includegraphics[height=6.5cm]{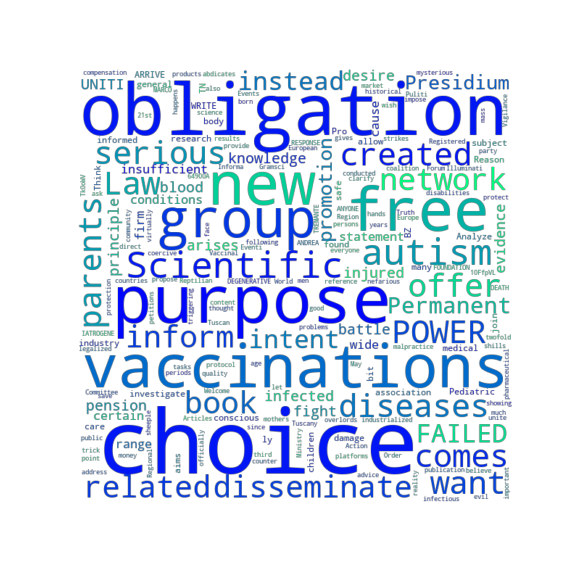}
\caption{\label{fig_wordcloud}Wordcloud showing the most used words by PV pages (left) and AV pages (right).}
\end{figure*}

To this extent, we present an exploratory visualisation of the words contained in the descriptions of the pages associated with each group (PV and AV), as a wordcloud. 
A wordcloud is a graphical representation of a ``cloud'' of words in which each word is sized according to its multiplicity in the document.
Figure~\ref{fig_wordcloud} depicts the most common words that appear in the description of the PV and AV groups of pages~\footnote{For the translation we have used the library Python Googletrans 2.4.0.}. 

In a glance, we observe that PV pages emphasise the concepts of children and parents, join and love, misinformation, unite. On the other hand, AV pages focus on the idea of free choice, purpose, obligation and power. 
This is in line with the study of Amin et al.~\cite{Amin2017} who stated that vaccine-hesitant individuals are emphasising more the moral value of ``liberty'', the so-called sixth moral foundation.  The above items are expressing a similar notion to the ones included in the ``liberty foundation'' scale employed to quantify this scale \cite{iyer2012understanding}.
``Scientific'' and ``book'', also appears in the AV discourses, indicating that AV community is not unwilling to engage with scientific evidence, but perhaps there is a biased interpretation of experimental outcomes~\cite{browne2015going}.
Diving further into the moral rhetoric of the PV and AV page descriptions, we employed the dictionary proposed by Graham et al. \cite{graham2009liberals} That relates specific words to the five predefined dimensions defined at the MFT. We found that PV pages contain more words associated with the moral dimension of care and authority while AV pages focus more on the value of loyalty.  
A limitation of the dictionary though is that the words related to each trait are few and some of them too formal for the everyday communication used in social media.

\subsection{Facebook Page Embeddings}

Network embeddings are an expanding field~\cite{tang2015line}, increasingly employed in a variety of tasks, such as network structure identity and community detection  \cite{PerozziAS14,wang2016structural,ribeiro2017struc2vec,wang2016structural},
but also topics of social interest such as classification of demographics and other behavioural attributes \cite{Perozzi:2015:EAP:2740908.2742765,zhang2016homophily}, to social interactions and behaviour prediction \cite{cho2016latent}.
Here, we propose an embedding based on the co-occurrence of pages people ``Liked'' on Facebook.
Such vision of the network allows us to explore common interests that two individuals may have since when two nodes are physically closer in the network, the more Facebook pages they share (e.g. have commonly ``Liked'').
Initially, we employed this technique to pages related to ``health'', since from the previous analysis it emerged as a domain with a significant differentiation in the PV and AV cohorts.  

\begin{figure}
\centering
\includegraphics[width = \columnwidth]{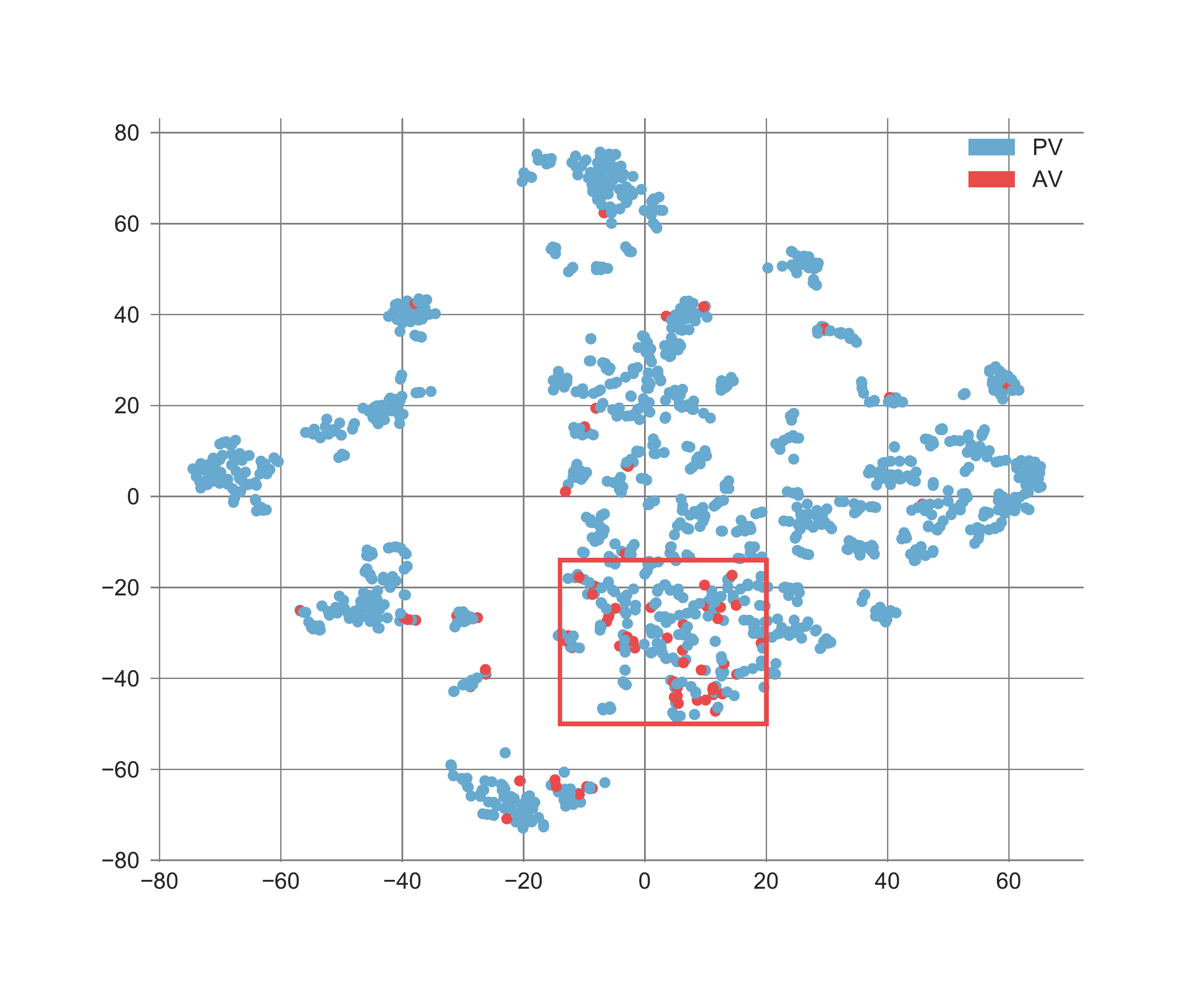}
\caption{\label{health_emb}Health co-occurence network embedding with t-SNE visualization. Embedding parameters: dimension=50, walk length = 5, number of walks = 50, p=2,q=0.8}
\end{figure}

We perform parameter tuning with grid search of \textit{p} and \textit{q} in the set \{4,2,1,0.8,0.1\}, always keeping $q<1$ since we are interested in the community structure of the network and the notion of homophily. 
We visualise the embedding in two dimensions employing the t-SNE algorithm \cite{maaten2008visualizing}; Figure~\ref{health_emb} depicts the network generated by the likes in the pages related to health topics. 
Zooming in the area with the highest concentration of AV participants, we retrieve the Facebook pages commonly ``Liked'' by these participants (Table~\ref{health_pages_area}). 
We observe that many of the emerging pages are related to food and nutrition. In particular, participants in this area ``Like'' pages about complementary and alternative medicine like for instance, ``Apoteca Natura'' (Natural Pharmacy), ``Cure Naturali'' (Natural Cures), and ``Omnama'' (spiritual consultancy page),  as well as medical doctors who support anti-vaccination attitudes (``Dr. Roberto Gava'').
From the emerging topics, we see that these interests are in line with the claims that negative attitudes to vaccination are related to the preferences of alternative medicine, as well as the endorsement of spirituality \cite{browne2015going}.
This alternative view of the network, allows us to explore interests that people with shared views on a specific topic may share in a data-driven way.

\begin{table}
\begin{tabular}{ll}
\hline
Facebook Page & Total Likes \\%among participants \\
\hline
Psychology (Psicologia)&19\\
The delicious good eating & 17\\
(Il Goloso Mangiar Sano) & \\
VaccinYES (VaccinarSI)&16\\
Omnama & 12\\
Self-weaning (Autosvezzamento)& 12\\
Natural Cures (Cure Naturali)&11\\
You know health and nutrition & 10\\
(Lo Sai Salute e Alimentazione)& \\
San Donato Hospital Group &9\\
(Gruppo Ospedaliero San Donato)&\\
Delicious Magazines (Buonissimo Magazine)&9\\
FruitsWeb (FruttaWeb)& 9\\
wellMe.it &8\\
Rita Levi Montalcini& 8\\
Dr. Roberto Gava &8\\
Natural Pharmacy (Apoteca Natura)&6\\
\hline
\end{tabular}
\caption{The 15 most liked Facebook Pages by participants in the area with the highest concentration of AV supporters, see Figure~\ref{health_emb}, red box.}
\label{health_pages_area}
\end{table}

\section{Limitations}

There are several limitations to our work. 
First and foremost, population in our cohort that has filled in questionnaires, especially the ones about morals and personality is small, affecting the statistical significance of our results substantially.
Also, many people interested in vaccine controversies are not necessarily following a Facebook page on the argument; hence, do not result in our analysis.

\section{Conclusions and Future Work}

Decision-making processes are complex; multiple factors contribute to how people rationalise, form opinions, and finally take actions.
When it comes to critical societal issues such as vaccine hesitancy, it is essential to understand the driver  of this process, aiming at bridging the gap between the two sides.
The determinants of attitude formation towards vaccination may range from psychological, moral, cultural, or even societal.
%and may contribute to negative vaccination attitudes 
\cite{bean2011emerging,kata2010postmodern,yaqub2014attitudes}.

%Social media increased the potential for communication and diffusion of information, however, opinion polarisation and formation of echo chambers in the discussion of several topics, such as vaccination, make public debates more difficult~\cite{kata2010postmodern,dunn2015associations, dredze2016zika,schmidt2018polarization}.

In this study, we analysed the psychological, moral, and political views of a large cohort in Italy, employing surveys administered on a Facebook-hosted application. We assumed that  ``liking'' a Page that expresses a positive or negative attitudes towards vaccination, the participant shares the same views on the topic.
%Our findings are in line with the current results in the literature, even if our sample size is limited. 

Analysing people's political views, we found those who are sceptical towards vaccination trust less the legal norms, both in a national and at a European level. Contrary to what is reported by a study carried in the US~\cite{Amin2017}, vaccine hesitancy is related to a tendency to argue that newer lifestyles contribute to the decline of the society while feeling the urge to defend more the traditional religious and moral values. 
%Supporters of vaccination, instead, have a more favourable opinion regarding the immigrants and their assimilation in the society. 
%Taking into consideration the moral values, those with contrary opinions towards vaccination appear to be more individualists, meaning that they focus more on the protection and fair treatment of the individuals, contrary to those who accept vaccination, who seem to emphasise more the social binding foundation, putting the society as a priority over the individual. Although our cohort is small, these findings are in line with~\cite{Amin2017}.
The personality determinants of vaccine hesitancy relate to low agreeableness and high openness to experience traits.
Regarding general interests, vaccine-hesitant individuals appeared to be more interested in topics like sports, hobbies and health.

Moving from surveys to digital data, we focused on linguistic usage employed to express the moral narratives, in the description of the Facebook pages, of both communities (AV and PV).
In line with the moral values expressed above, the narratives of supporters of vaccination and pro-vaccine campaigns emphasise more in notions of care and protection, while the opposite side emphasises more on the value of freedom of choice, science,  power, and law, which relate to authority.
Finally, from the network embedding based on the Pages ``Liked'' related to ``health'', we find pages related to alternative medicine and spirituality.
This alternative view of the network, allows us to explore interests that people with shared views on a specific topic may share in a data-driven way. 

As Debe et al.~\cite{dube2015strategies} suggest, due to the multi-facet nature of vaccine hesitancy there is still no effective communication strategy.
%People with a particular moral profile can react differently depending on the way a specific message is framed, for example, in political influence~\cite{feinberg15}. 
%Similarly, the use of specific linguistic cues in argumentation might trigger moral values persuasion~\cite{yang2018visual}. 
Here, we propose an approach where a social media platform is employed as an alternative tool to help researchers and policymakers better understand the psychological and moral views of vaccine-hesitant individuals.
%adding additional value to the usage of social marketing practices in immunisation programs~\cite{carroll2002public, nowak2015addressing}.
%The broader goal of this study is to explore the potentials of a social media platform such as Facebook to act as a sensing tool, both at an individual but also at a community level.
%Such insights may lead to communication campaigns that are more likely to appeal to the reasoning criteria of vaccine-hesitant people.

Future steps include, first of all, the expansion of the cohort. Secondly, in this study we only assessed the linguistic content of the page descriptions; we plan to expand this task in the actual content of the pages since this will provide a more unobstructed view on the arguments made by both sides.

In conclusion, insights from social media data may inform interventions that take into consideration the cognitive, moral, psychological, and political values of people hesitant towards vaccination, building general confidence by appealing to their values.

\begin{acks}
The authors acknowledge support from the ``Lagrange Project'' of the ISI Foundation funded by the Fondazione CRT. 
They would like to thank all the participants in the cohort.
\end{acks}

%
% The next two lines define the bibliography style to be used, and the bibliography file.
\bibliographystyle{ACM-Reference-Format}
\balance
%\bibliography{sample.bib,DemographicResearch.bib,MFTClassification.bib,Embedding.bib}

%%% -*-BibTeX-*-
%%% Do NOT edit. File created by BibTeX with style
%%% ACM-Reference-Format-Journals [18-Jan-2012].

\end{document}